# Optical, luminescence, and scintillation properties of advanced ZnWO4 crystal scintillators

P. Belli [a,b], R. Bernabei [a,b,2], Yu.A. Borovlev [c,d], F. Cappella [e,f], V. Caracciolo [a,b], R. Cerulli [a,b], F.A. Danevich [g], V.Ya. Degoda [h,], A. Incicchitti [e,f], D.V. Kasperovych [g], Ya.P. Kogut [h], A. Leoncini [a,b], G.P. Podust [h], A.G. Postupaeva [d], V.N. Shlegel [c]

[a] INFN, sezione di Roma "Tor Vergata", I-00133 Rome, Italy
[b] Dipartimento di Fisica, Università di Roma "Tor Vergata", I-00133 Rome, Italy
[c] Nikolaev Institute of Inorganic Chemistry, 630090 Novosibirsk, Russia
[d] CML Ltd, 630090 Novosibirsk, Russia
[e] INFN, sezione di Roma, I-00185 Rome, Italy
[f] Dipartimento di Fisica, Università di Roma "La Sapienza", I-00185 Rome, Italy
[g] Institute for Nuclear Research of NASU, 03028 Kyiv, Ukraine
[h] Taras Shevchenko National University of Kyiv, 01601 Kyiv, Ukraine

**Abstract**

Zinc tungstate (ZnWO4) crystal scintillators are promising detection material for the experiments searching for double beta decay, dark matter, and investigating rare alpha decays. An extended R&D was performed to develop advanced quality ZnWO4 crystal scintillators. The R&D programme included the selection of the initial materials, the variation of the compound stoichiometry, the application of single and double crystallization, and the annealing of the crystal boules. The optical transmittance of the produced boules was measured, and the luminescence under X-ray excitation in the temperature region from 85 K to room temperature was studied (thermally stimulated luminescence was measured till 350 K). The energy resolution and the relative scintillation pulse amplitude were measured with gamma-sources demonstrating high scintillation properties of the samples produced by single crystallization from deeply purified zinc and tungsten oxides, with stoichiometric composition, annealed in air atmosphere.

## 1. Introduction

The zinc tungstate (ZnWO4) is a crystal luminescent material well-known since more than seventy years [1]. The ZnWO4 crystal growth techniques were developed, and the luminescence and the scintillation properties of the material were studied taking into account its high stopping power and high scintillation efficiency [2, 3, 4, 5, 6, 7, 8, 9]. Effects of doping by different elements to increase ZnWO4 crystals light yield were investigated too [10, 11, 12, 13, 14, 15, 16, 17].

The material was considered for the first time in the 1980s as possible detector to search for double beta decay of Zn and W isotopes [18]. The first low-background measurements with a small 4.5 g scintillator were performed at the Solotvina Underground Laboratory to demonstrate applicability of ZnWO4 crystal scintillators for double beta decay and dark matter experiments [19]. Further development of the material was also motivated by attempts for increasing the crystals volume, for improving the optical and the scintillation characteristics [20, 21, 22] and for utilizing as low-temperature scintillation bolometer [23, 24]. Consequently, high sensitivity double beta decay experiments were realized by using large volume and high optical quality $ZnWO_4$ crystal scintillators [25, 26, 27]. The results were obtained thanks to the very high level of radiopurity of the material [28, 29, 30].

---

[2] e-mail: rita.bernabei@roma2.infn.it (corresponding author)

Anisotropic properties of the $ZnWO_4$ scintillators offer an interesting possibility to exploit the directionality approach to investigate the presence of dark matter candidates that induce nuclear recoils [31, 32]. The goal of the ADAMO project is the development of $ZnWO_4$ crystal scintillators with extremely low radioactive contamination and very high optical and scintillation properties capable to provide as low as possible energy threshold and background level at low energies. The R&D of advanced large volume $ZnWO_4$ crystal scintillators for low-counting experiments is in progress [33].

The development of high scintillation efficiency $ZnWO_4$ scintillators requires a clear understanding of the scintillation process, emission nature and mechanisms, study of the production technology, quality, and composition of the raw materials for the crystals' growth on the optical and scintillation properties of the material. The production of several samples of $ZnWO_4$ crystals will be described in the next Section. Experimental techniques, measurements, results, and discussion of optical, luminescence and scintillation measurements are presented in Section 3. The main results of the study are summarised in the Conclusions Section.

## 2. Development of advanced $ZnWO_4$ crystal scintillators

High quality large volume $ZnWO_4$ crystal scintillators have been developed in the Nikolaev Institute of Inorganic Chemistry (NIIC, Novosibirsk, Russia) with the help of the low-thermal gradient Czochralski technique (LTG Cz) [33, 34, 35, 36, 37]. The samples studied in the present work were produced by the Crystal Manufacturing Laboratory Ltd. (CML, Novosibirsk, Russia) in air atmosphere in a platinum crucible 50 mm in diameter and 100 mm height installed into three zone resistance heater. The ratio between ZnO and $WO_3$ in the initial powder for the crystal growth was chosen according to the phase diagram of the $ZnO-WO_3$ system. The powder was synthesized by using reaction $WO_3 + ZnO \rightarrow ZnWO_4$. High purity grade zinc oxide (99.995%) produced by Umicore (Belgium) was used for the R&D. Several samples of tungsten oxide of different origin were utilized:

- synthesized at the NIIC with Si concentration <50 ppm and concentration of transition metals no more than 1 ppm, the material development is described in [35] (denoted hereafter as "$WO_3$ NIIC I");
- purified at the NIIC using an additional process of sublimation of tungsten chlorides (denoted as "$WO_3$ NIIC II");
- manufactured by Nippon Tungsten Co., Ltd. (Japan);
- manufactured by Japan New Metals Co., Ltd (JNM, 4N grade CWO) with a maximum Fe content <1 ppm and a maximum Mo content less than 10 ppm.

The rotation speed during the crystal growth was in the range from 3 rpm to 6 rpm, the crystallization rate was 1.5 mm / h, all crystals were grown in the [010] direction by using oriented crystal seeds. The single crystallization samples were grown from the initial powder, while the double crystallization samples were grown by recrystallization of the single crystallization crystalline material. In addition to the crystals produced from stoichiometric $ZnWO_4$ compound, some amount of ZnO or WO₃ (see Table 1) was added to study dependence of the material properties on its composition. After the crystal growth process, the crystal boules were annealed in air atmosphere over 24 h (except the boules produced by double crystallization, see Table 1).

Table 1. The samples of $ZnWO_4$ crystals used in the present study and the boules of origin.

| Crystal boule | Sample size (mm³) | Number of crystallizations | WO₃ | Compound stoichiometry |
|---|---|---|---|---|
| No. 75 | $10 \times 10 \times 2$ $\varnothing 30 \times 60$ | Double | NIIC II | + 0.3 % of WO₃ |
| No. 76 | $10 \times 10 \times 2$ $\varnothing 30 \times 60$ | Double | Nippon Tungsten Co., Ltd | + 0.25 % of ZnO |
| No. 83 | $10 \times 10 \times 2$ $\varnothing 30 \times 60$ | Single, annealed | NIIC I | + 0.15 % of WO₃ |
| No. 84 | $10 \times 10 \times 2$ $\varnothing 30 \times 60$ | Single, annealed | NIIC I | stoichiometric |
| No. 85 | $10 \times 10 \times 2$ $\varnothing 30 \times 60$ | Single, annealed | Japan New Metals Co., Ltd | stoichiometric |
| No. 91 | $\varnothing 30 \times 67$ | Single, annealed | NIIC I | stoichiometric |
| No. 94 | $\varnothing 30 \times 31$ $\varnothing 30 \times 32$ | Single, annealed | NIIC I | stoichiometric |

The crystalline boules are shown in Fig. 1. One can see that, while the boules No. 75 and 76 have the typical pink colour of $ZnWO_4$ crystals, the optical quality of the crystals No. 83–85 and 91 is rather high.

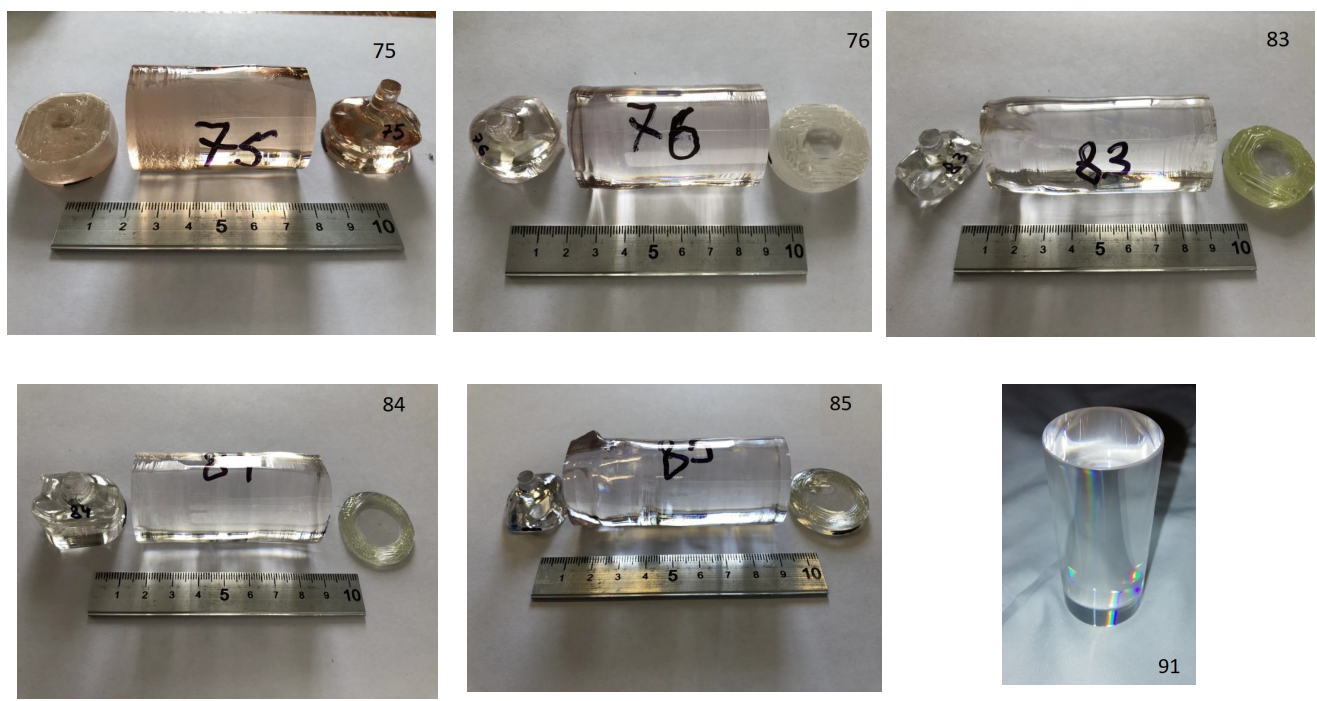

Fig. 1. Photographs of the crystalline boules 75, 76, 83−85 and sample with sizes $\varnothing 30 \times 67$ mm$^3$ produced from the crystalline boule No. 91 used in the present study. The crystal boules numbers are given on the photographs.

Samples with sizes $\varnothing 30 \times 60$ mm$^3$ with optically polished faces were cut from the boules for optical transmission measurements. One more sample for optical measurements was cut from the boule No. 91 with sizes $\varnothing 30 \times 67$ mm$^3$ (the sample is shown in Fig. 1 too). Samples with sizes $10 \times 10 \times 2$ mm$^3$ were cut from the boules for the luminescence investigations. The samples surfaces were optically polished. Two samples of cylindrical shape with sizes $\varnothing 30 \times 31$ mm$^3$ and $\varnothing 30 \times 32$ mm$^3$ were produced from the boule No. 94. One was used for energy resolution, and both are now used for low-background tests (the low-background measurements are in progress and will be described in a separate report). All the samples are listed in Table 1.

## 3. Measurements, results, and discussion

### 3.1. Optical transmission

The optical transmission of the ZnWO$_4$ cylindrical-shaped samples with sizes $\varnothing 30 \times 60$ mm$^3$ produced from the crystal boules No. 75, 76, 83−85 was measured by using a UV-2201 Double-beam UV-spectrophotometer from Shimadzu. The face surfaces of the samples for the transmission measurements were optically polished. In order to compensate the light losses at the surfaces of the samples due to the Fresnel reflection caused by the high refraction index (that depends on the wavelength), a 1-mm-thick polished ZnWO$_4$ disk was installed in the reference channel of the spectrophotometer. The optical transmission of the ZnWO$_4$ sample No. 91 was measured using a Perkin Elmer UV/VIS spectrometer Lambda 18. A thin (1.8 mm) sample of the ZnWO$_4$ crystal was placed in the reference beam of the instrument to correct the reflection losses. The results of the measurements are presented in Fig. 2. In general, the transmission spectra agree with the literature data [14, 20, 21, 23, 38]. However, the transmission varies substantially depending on the sample production protocol. In particular, the samples produced by double crystallization are definitely of worse optical quality.

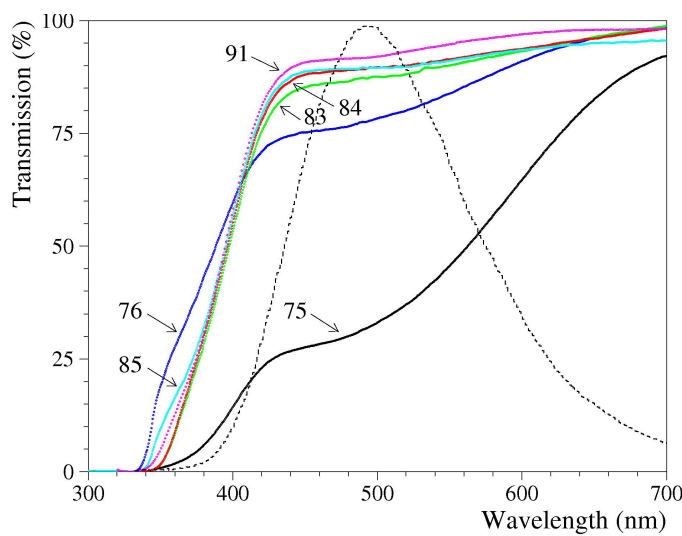

Fig. 2. Optical transmission spectra of single (samples 83–85, 91) and double crystallization (samples 75 and 76) ZnWO₄ crystals. Emission spectrum of one of the samples at room temperature is shown by dashed line (see Section 3.2).

### 3.2. Luminescence under X-ray excitation

#### 3.2.1. Experimental

To measure the luminescence of ZnWO₄ crystals under X-ray excitation (XRL) the samples with sizes $10 \times 10 \times 2$ mm³ were installed on a copper holder inside a vacuum cryostat and irradiated by X-ray through a beryllium window. The temperature of the crystal holder was controlled by using a chromel-copel thermocouple and a semiconductor silicon sensor. An X-ray tube with a copper anode operated at 20 kV, with current 25 mA and flux of 0.25 mW/cm² was used as an X-ray source. XRL of the ZnWO₄ samples was measured in a wide region where the spectral sensitivity of the set-up was determined by the used photomultiplier tube (PMT, FEU-106, sensitive in a wide wavelength region of 350−820 nm). The emission spectra were registered by a second PMT FEU-106 after selection of a narrow spectral range by using a monochromator MDR-2 with a 600 mm⁻¹ diffraction grating.

#### 3.2.2. Emission spectra

The emission spectra of the ZnWO₄ crystal samples under X-ray irradiation at 85 K and 295 K, corrected for the spectral sensitivity of the set-up, are shown in Fig. 3. The spectra are peaked at ≈494 nm at room temperature and at ≈502 nm at 85 K. The results are in a reasonable agreement with the reported values 479-502 nm for ZnWO₄ crystals under X-ray excitation [3, 4, 14, 15, 39, 40]. A certain shift of the emission spectral maximum is observed in all the samples. One can see a clear difference in the XRL intensity of the samples[3]. The spread of the XRL intensity is bigger at room temperature. The difference can be explained by presence of nonradiative recombination centres

---

(caused by defects of different nature) that compete with the radiative recombination centres. It is natural to assume that the concentration of defects depends on the crystals production technology.

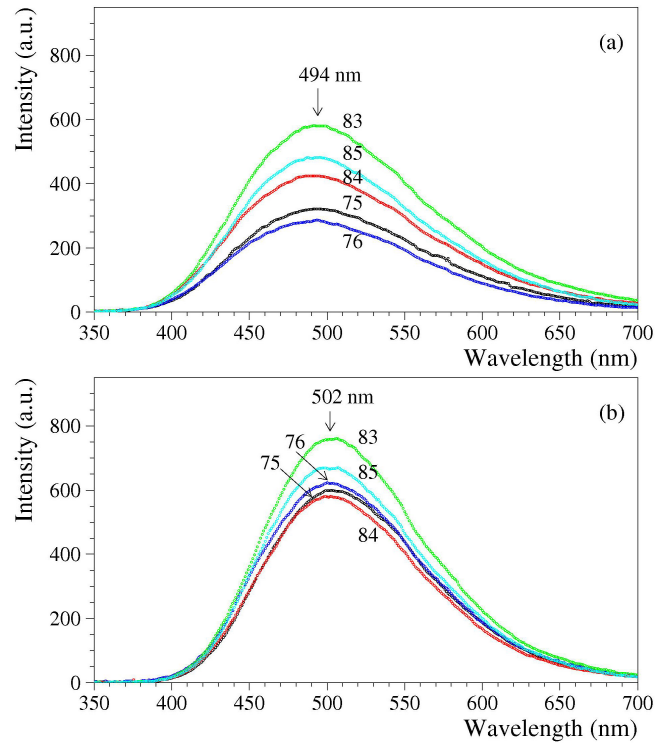

Fig. 3. Emission spectra of the ZnWO₄ samples under X-ray irradiation at 295 K (a) and 85 K (b). The labels "75", "76", "83", "84", "85" correspond to the sample numbers.

### 3.2.3. Temperature and dose dependence of XRL intensity

The dependence of the ZnWO₄ samples XRL intensity on temperature was measured at the sample cooling from (278 – 293) K to ≈ 90 K. The dependence is shown in Fig. 4. The luminescence intensity increased with temperature decrease, in agreement with the data of other studies [40, 41]. It should be stressed that increase of the scintillation signals amplitude with decrease of temperature is higher, roughly by a factor 2 [5, 22, 23]. It can be explained by different nature of scintillation and luminescence response (see also Fig. 9 and related discussion in Section 3.3).

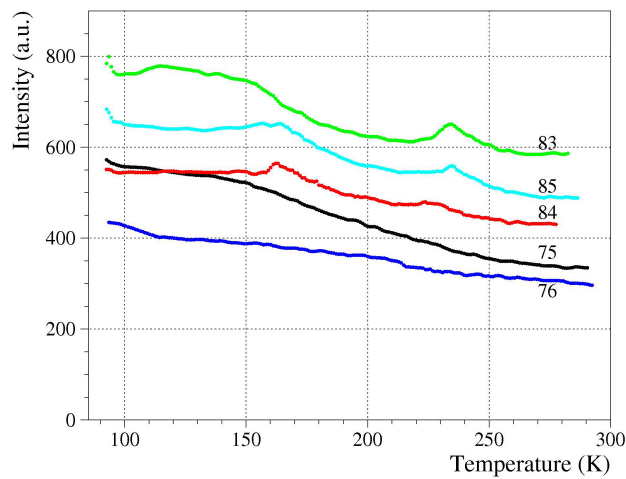

Fig. 4. Temperature dependency of the ZnWO₄ crystal samples XRL intensity. The curves variations can be explained by uneven cooling rate of the samples.

The dose dependences of the XRL intensity was measured under constant intensity X-ray radiation. The XRL intensity decreases with accumulation of dose (see Fig. 5). A decrease of a ZnWO$_4$ scintillator light output under X-ray irradiation was also reported in [42].

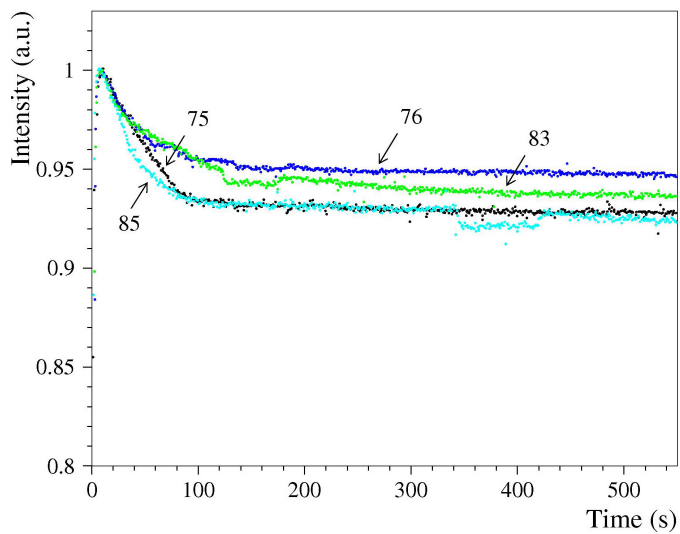

Fig. 5. Dose dependence of the ZnWO$_4$ samples XRL intensity measured at room temperature. The data are normalised on the maxima values 7–9 sec after measurement start. The jumps on the curves are due to unstable operation of the X-ray tube.

### 3.2.4. Phosphorescence

The ZnWO$_4$ samples phosphorescence have been measured for ≈8 min after irradiation of 20 min at 85 K with an exposure dose of 0.30 J/cm$^2$. Rather intensive slowly decaying phosphorescence component is observed after X-ray irradiation at 85 K (see Fig. 6). The observation indicates a noticeable number of different traps in the ZnWO$_4$ crystal samples. It should be noted that the sensitivity of the recording system to register the phosphorescence was increased by two orders of magnitude. This means that the concentration of the traps that cause phosphorescence are orders of magnitude lower than the concentration of the radiative recombination centres in the ZnWO$_4$ crystals.

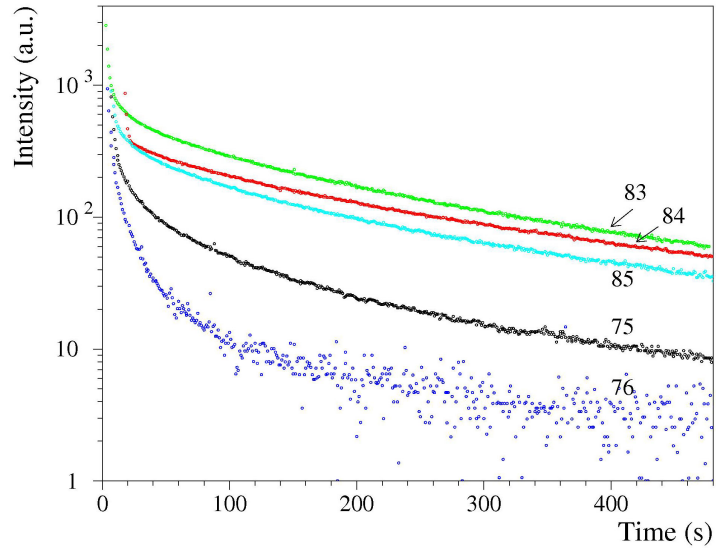

Fig. 6. XRL phosphorescence of the ZnWO₄ samples measured at 85 K after X-ray irradiation over 20 min.

### 3.2.5. Thermally stimulated luminescence

After the phosphorescence measurements the samples were heated with a rate $(0.30 \pm 0.02)$ K/s to study the thermally stimulated luminescence (TSL). TSL was observed in the ZnWO₄ crystal samples after irradiation by high energy γ-rays at low temperature [17, 14]. In the present study TSL was registered after X-ray irradiation over 20 min (that corresponds to a dose 0.30 J/cm²) at 85 K. The TSL curves are presented in Fig. 7. It should be noted that no detectable TSL and phosphorescence were observed after the sample irradiation at 295 K.

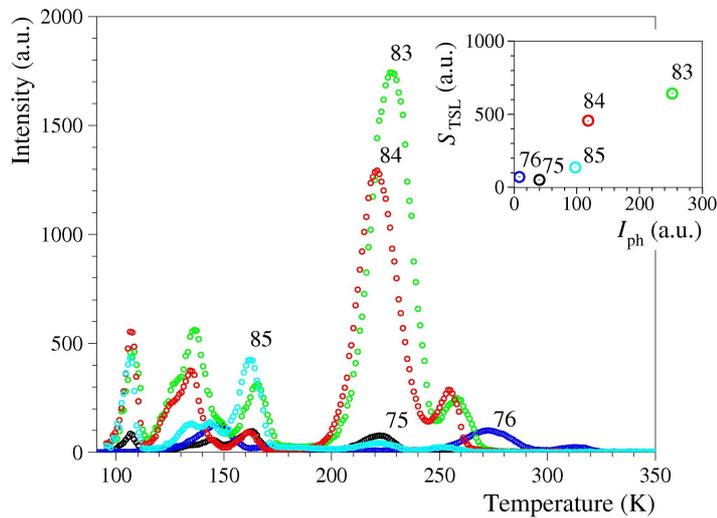

Fig. 7. Thermally stimulated luminescence of ZnWO₄ crystal measured after X-ray irradiation over 20 min at temperature 85 K. Inset: area of TSL curves in the temperature interval 95-320 C ($S_{TSL}$) versus phosphorescence intensity at 200 sec after the X-ray irradiation termination ($I_{ph}$).

From the TSL data one can conclude that all the crystal samples have similar set of traps, however with different concentration. For instance, the concentration of traps in the samples No. 83, 84 and 85 (produced by single crystallization) is much higher than in the samples No. 75 and 76 (grown by double crystallization).

Areas of the TSL curves in the temperature interval 95-320 C ($S_{TSL}$) versus phosphorescence intensities at 200 sec after the X-ray irradiation termination ($I_{ph}$) are presented in Inset of Fig. 7. It should be stressed that phosphorescence and TSL appear due to different traps. Thus, some correlation between the TSL and phosphorescence intensities indicates that the concentration of all kinds of traps in the crystal varies depending on the growing technology.

### 3.3. Scintillation properties

Scintillation measurements of the $10 \times 10 \times 2$ mm samples produced from the crystal boules No. 75, 76, 83 − 85 were performed by using a 3 in. photomultiplier tube Hamamatsu model R6233-100 with a bialkali photocathode with a spectral response in the wavelength interval from 300 nm to 650 nm, with a peak sensitivity at 420 nm. The crystal samples were optically coupled to the PMTs photocathode with the help of a Dow Corning Q2-3067 optical couplant and covered by Teflon tape to improve the scintillation light collection. The signals from the PMT entered a SAMP-04 spectroscopy amplifier with a 15 μs shaping time and then were processed by a peak sensitive analog-to-digital converter (both units produced by the Intelligent Electronic Systems, Kharkiv, Ukraine). The crystals surface was diffused by sanding paper with an average grain size 20 μm. The energy spectrum of γ-ray quanta of $^{137}$Cs, $^{207}$Bi, and $^{241}$Am γ-sources measured with the ZnWO$_4$ crystal sample No. 84 (showing the highest light output) is presented in Fig. 8. It should be stressed that so high energy resolution ($R$, full-width half-maximum over peak position) was never reported for ZnWO$_4$ crystal scintillators.

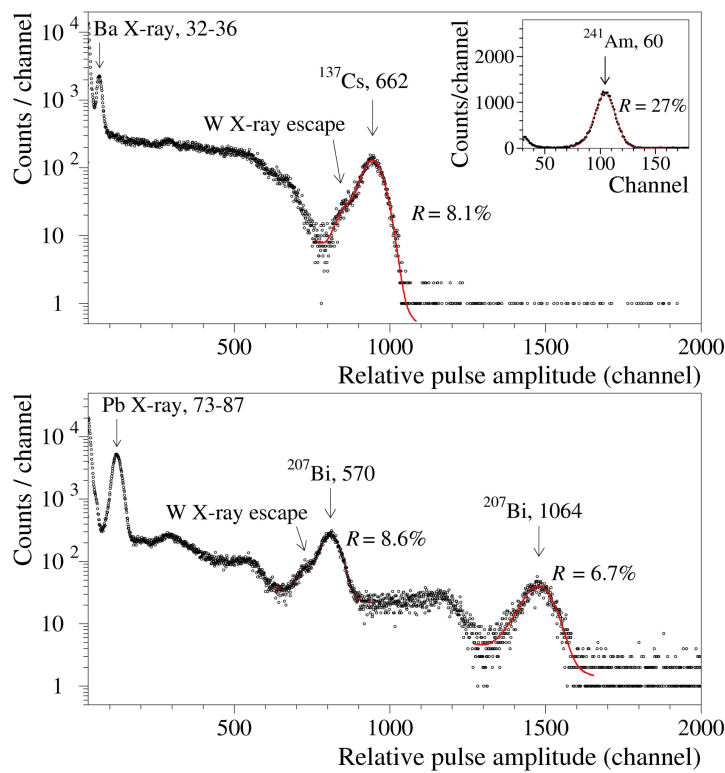

Fig. 8. The energy spectra of γ-ray quanta of $^{137}$Cs and $^{207}$Bi measured by scintillation detector with the ZnWO$_4$ crystal sample No. 84. Energy spectrum of γ-ray quanta of $^{241}$Am is shown in Inset. Energy of X-ray and γ-ray quanta is in keV.

The data on the relative scintillation pulse amplitude (estimated from the scintillation measurements for 570-keV γ-peak of [207]Bi) versus relative luminescence intensity of the ZnWO$_4$ crystal samples at room temperature are presented at Fig. 9 (filled circles). These two characteristics do not correlate as one naively would expect. Possible explanations can be a contribution of phosphorescence that does not affect the scintillation pulse amplitude, and a dose dependence of XRL intensity that is negligible in scintillation measurements.

To estimate a possible degradation of ZnWO$_4$ scintillation efficiency after the X-ray irradiation, the samples that have not been used for the XRL measurements were investigated as scintillators in the same conditions (see Fig. 9, open circles). Despite the XRL measurements were performed in December 2020, one year before the scintillation test (in December 2021), the samples irradiated with a dose $\sim (1.4-1.6)$ W $\times$ cm$^2$ showed a slightly lower scintillation light output at level of $1.4\% - 5.7\%$ than the non-irradiated ones.

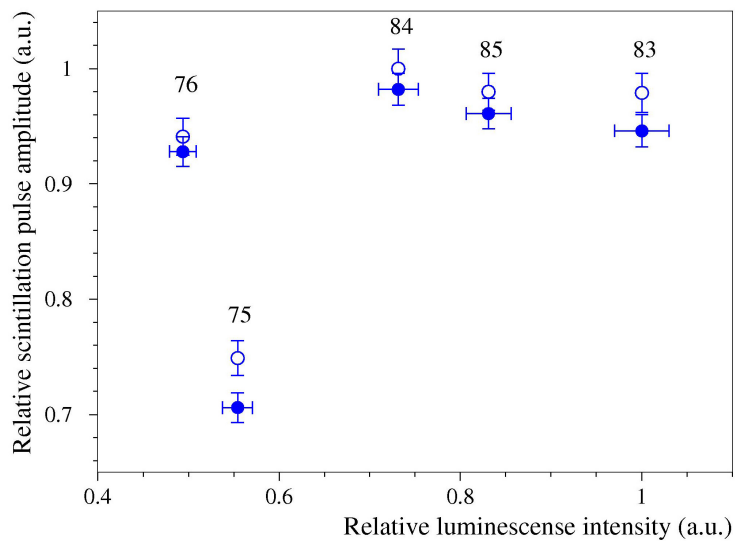

Fig. 9. Relative scintillation pulse amplitude versus relative XRL intensity of the ZnWO$_4$ crystal samples at room temperature (filled circles). Results of scintillation measurements with non-irradiated samples are shown by open circles.

The scintillation properties of a ZnWO$_4$ scintillator $\varnothing 30 \times 31$ mm$^3$ cut from the boule No. 94 were tested with the same equipment. The crystals surface was diffused by sanding paper with an average grain size 46 μm. For scintillation light collection the scintillator was wrapped by few layers of Teflon tape and by one layer of aluminized film. Energy spectra measured by the scintillation detector with [60]Co, [137]Cs, [232]Th and [241]Am γ-sources are presented in Fig. 10. The obtained energy resolutions are rather high; they are near the best reported for ZnWO$_4$ crystal scintillators [20, 22, 23, 43, 44].

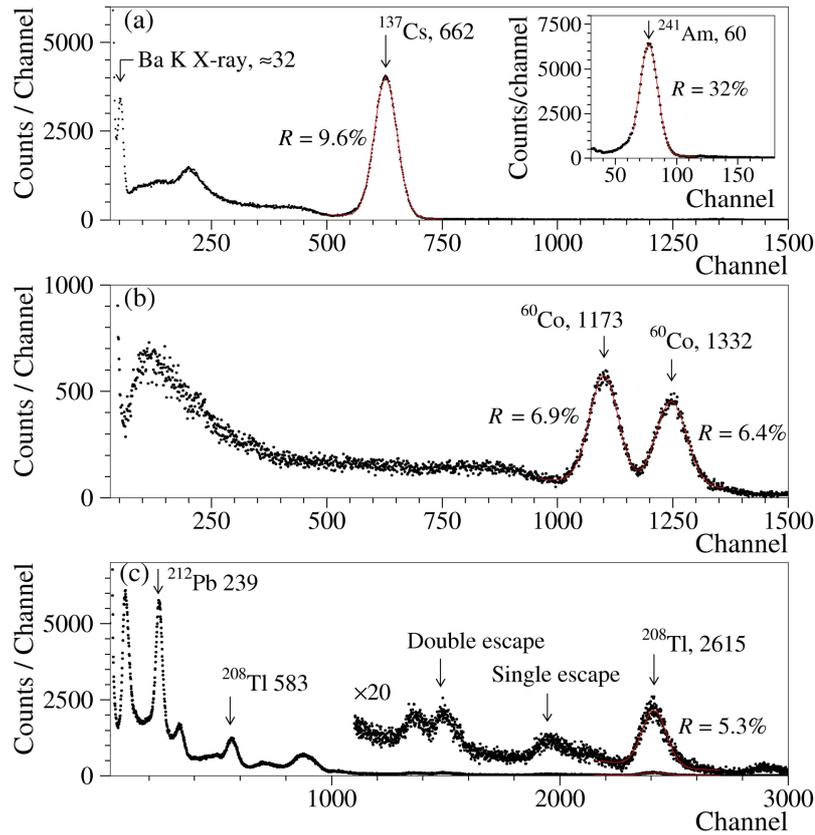

Fig. 10. Energy spectra of γ-ray quanta of $^{137}$Cs (a), $^{60}$Co (b) and $^{232}$Th (c) measured by scintillation detector with the ZnWO$_4$ crystal No. 94 with sizes ⌀30×31 mm. Energy spectrum of $^{241}$Am is shown in Inset.

## 5. Conclusions

High optical and scintillation properties ZnWO$_4$ crystal scintillators were developed by using the low-thermal gradient Czochralski technique after an extended R&D that included variation of the compound stoichiometry, using of initial WO$_3$ of different producers and additionally purified, utilization of single and double crystallization with and without annealing of the grown boules.

Luminescence under X-ray excitation of the developed ZnWO$_4$ crystals (emission spectra, temperature and dose dependencies of the luminescence intensity, phosphorescence, and thermally stimulated luminescence) was studied in the temperature region from 85 K to room temperature (thermally stimulated luminescence was measured till 350 K). Scintillation properties of the ZnWO$_4$ crystals were tested with $^{60}$Co, $^{137}$Cs, $^{207}$Bi, $^{232}$Th and $^{241}$Am γ-ray sources, optical transmission spectra of the samples were measured in the wavelength interval 300–700 nm. The best optical and scintillation characteristics were obtained with ZnWO$_4$ crystal samples grown by single crystallization from the ZnWO$_4$ compound of the stoichiometric composition prepared from deeply purified WO$_3$, annealed in air atmosphere. Surprisingly the double crystallization leads to deterioration of the optical, luminescent and scintillation properties of the material, despite lower level of phosphorescence and thermally stimulated luminescence that is evidence of a lower defects' concentration. It can be explained by the increase of competing centres of nonradiative recombination concentration (further study of the material luminescence mechanisms is a subject of another report [45]). No clear correlation was observed between the scintillation light output and luminescence intensity of the samples that can be explained by phosphorescence and dose dependence of XRL intensity that are both negligible in scintillation measurements. The absence of correlation between the luminescence intensity and the scintillation pulse amplitude indicates that the achieved scintillators quality (especially of the samples produced by double crystallization) is not perfect and there is still room to improve the $ZnWO_4$ production technology.

Low-background measurements of two highest optical quality crystal samples produced from the crystal boule No. 94 are running at the INFN Gran Sasso underground laboratory to estimate radioactive contamination of the material. A further R&D is in progress aiming at development of larger volume crystals for low counting experiments.

## Acknowledgements

Yu.A. Borovlev and V.N. Shlegel were supported by the Ministry of Science and Higher Education of the Russian Federation N121031700314-5. F.A.Danevich and D.V. Kasperovych were supported in part by the National Research Foundation of Ukraine Grant No. 2020.02/0011.